\documentclass[10pt]{iopart}
\usepackage{iopams} 
\usepackage[]{epsfig}
\begin{document}

\title[Yield stress, heterogeneities
and activated processes in soft glassy materials]{Yield stress, heterogeneities
and activated processes in soft glassy materials}

\author{Ludovic Berthier}
\address{Theoretical Physics, 1 Keble Road, Oxford, OX1 3NP, UK}

\begin{abstract}
The rheological behavior 
of soft glassy materials basically 
results from the interplay between shearing forces and  
an intrinsic slow dynamics.
This competition can be described by a microscopic theory,
which can be viewed as a nonequilibrium schematic mode-coupling theory.
This statistical mechanics approach to rheology 
results in a series of detailed theoretical predictions, 
some of which still awaiting for their experimental verification.
We present new, preliminary, results about 
the description of yield stress, flow heterogeneities 
and activated processes within this theoretical framework.
\end{abstract}


\section{Introduction}

The discussion of `glassy' materials in textbooks is usually restricted
to simple molecular glasses, such as silica or polymeric glasses, 
and hence covers the standard field of the `glass transition'.
An impressive number of experiments performed in the last decade shows that
typical glassy effects are not restricted to structural glasses, but are
observed in a much wider variety of experimental systems. 
As a result, `glassy dynamics' is being actively 
studied in systems as different as dirty type II
superconductors, disordered ferromagnets or ferroelectrics, 
disordered electronic systems, soft glassy materials,
granular matter. These observations not only 
allow to draw interesting analogies between different systems, 
but also to use similar theoretical paths to describe various fields. 
This paper is concerned with the application of a theory initially developed 
in the context of the statistical mechanics of glasses to 
describe the rheology of soft glassy materials. 

The term `soft glassy materials'
was proposed in Ref.~\cite{sollich1} as a generic name for a large family of
complex fluids which share a similar phenomenology:
colloids, emulsions, pastes, clays...
It was then assumed
that their physical behavior basically results from the competition between 
an intrinsic glassiness, in the sense of large relaxation times,
and shearing forces which `accelerate' their dynamics.
This competition was then theoretically described using the simple
trap model of Ref.~\cite{trapmodel}, 
phenomenologically extended to account for 
the effect of an external flow. The model was further studied in 
Refs.~\cite{sollich2}. 
Since then, various phenomenological approaches have been 
proposed, which replace the concept of `traps' by
`fluidity'~\cite{derec}, `free volume'~\cite{lemaitre}, 
or `degree of jamming'~\cite{coussot}, 
but end up with extremely similar
mathematical formulations. 
The interest of such models is that they are simple enough so that
complicated flow and thermal histories can be easily implemented. 
The obvious drawback is that the physics is put by hand from 
the beginning: shear is assumed to reduce the trap depth, or to increase
the fluidity, the free volume, the degree of jamming.

The approach we discuss here is less transparent but 
does not assume anything beyond the dynamical evolution of a Langevin
type for a glassy system defined by its Hamiltonian~\cite{BBK}.
Interestingly, it can alternatively be viewed as a nonequilibrium extension
of schematic mode-coupling theories, as we describe in 
section~\ref{approach}.
A very similar approach was recently followed in Ref.~\cite{fuchs}.
Static and dynamic behaviors can be studied
in such detail that a very complete physical description can be proposed.
The theory is also sufficiently understood that its shortcomings, 
domain of validity, and, in principle, the path to possible improvements
are also known.
At present, only the case of steady shearing was investigated
in great detail, as reviewed in section~\ref{steady}.
We hope to report on transient behaviors in a near future,
the difficulties being mainly technical.
The paper also contains two new steps. 
First, we discuss the issue of a yield stress, 
and present preliminary results for its behavior
in section~\ref{yield}.
Second, we show in section~\ref{flow}
that two dynamical solutions for a given shear stress
are possible in a certain regime and discuss this feature in light of the 
recent observations of flow heterogeneities.

\section{A microscopic approach for nonlinear rheology}
\label{approach}

What would be an ideal theory for glassy rheology? Ideally, 
one would like to start from a microscopic equation 
for the dynamic evolution of the system under study, say a 
supercooled liquid in a shear flow, and solve this 
dynamics exactly. This `first-principles' approach is obviously an
extraordinary challenge and approximations have to be made
in order to get a system of closed dynamic equations. 
A well-known approximation in the field of the glass transition
is the mode-coupling approximation~\cite{bouchaud} which leads
to the mode-coupling theory (MCT) of the glass transition~\cite{gotze}. 
That the resulting equations can be obtained in a standard perturbative
way is discussed in Refs.~\cite{bouchaud,mazenko,kawasaki}, 
although the original derivation makes this less transparent. 

Generally speaking, starting from an evolution equation for the density
fluctuations
$\delta \rho(k,t)$, where $k$ is a wavector and $t$ is time, 
and a given pair potential interaction $V(k)$, 
the mode-coupling approximation amounts 
to a partial resummation of diagrams in the perturbative development
of the dynamical action.
One obtains dynamical equations (Dyson equations)
which close on two-point correlation $C(k,t,t') = \langle \delta \rho(k,t)
\delta \rho(-k,t') \rangle$ and conjugated response function
$\chi(k,t,t')$.
At thermal equilibrium, the fluctuation-dissipation 
theorem $T \chi(k,t,t') = \partial C(k,t,t') / \partial t'$,
implies coupled equations for the density 
correlators only: this is the MCT.
The theory has then the pair potential (or 
alternatively the static structure factor $S(k) = C(k,t,t)$) as an input, and
the dynamic behavior of the liquid, the correlators $C(k,t,t')$, as an 
output. 
`Schematic' models focusing on a given `important' wavevector, say
$k_0$, were formulated~\cite{schematic}.
This amounts to write $S(k) \approx 
1 + A \delta(k-k_0)$ and focus on $C(t,t') \equiv C(k_0,t,t')$ as a single 
observable. 
These schematic formulations lose the `ab-initio' character
of the full MCT, but they do essentially capture 
the dynamical singularities arising when the wave-vector 
dependence is kept~\cite{gotze}.

The MCT of the glass transition has been discussed 
at length in the literature~\cite{gotze,tarjus}. 
There are two points we would like
to emphasize, though.

(i) The `perturbative' derivation described above 
where both correlation and response are kept, opens
the door to the study of nonequilibrium behaviors.
The system can be out of equilibrium because the equilibration
time scale is too large for the experimental window: one 
focuses then on aging behaviors~\cite{reviewaging}.
The system can also be driven out of equilibrium by some 
external force, for instance a shear force.
This is the latter, rheological, situation we shall be interested in.

(ii) Kirkpatrick and Thirumalai remarked
that the schematic models can be exactly derived 
starting from some mean-field disordered 
Hamiltonian, like Potts and $p$-spin models~\cite{kithprl}.
The connection was studied further in 
an important series of papers~\cite{series}.
Beyond the basic connection made between two fields (spin and 
structural glasses), these works more importantly complement MCT
with all the knowledge one can get from the thermodynamic studies
of, say, the $p$-spin model.  
It is worth recalling that the $p$-spin model exactly 
realizes most of the thermodynamic 
`folklore' of the glass transition~\cite{tarjus}, such as 
such an entropy crisis, or the existence of 
many metastable states~\cite{kithprl}. 
Both sides, MCT and mean-field disordered models,
are now indissociable and form an ensemble that could generically be 
called the `mean-field theory of the glass transition'.
We shall see below how the knowledge 
of the free energy landscape (the `spin glass part' of the theory)
allows one to make qualitative predictions for the dynamics 
(the `MCT part'), 
that appears to be verified when the actual calculations are made.

\section{Steady rheology: A brief review}
\label{steady}

The microscopic approach to glassy rheology 
described in general terms in the previous section
was quantitatively investigated in Ref.~\cite{BBK}.
There, a schematic model (of the $F_p$ family~\cite{gotze}) was 
extended to take into account the crucial ingredient that
the dynamics is externally driven.
Technically, this amounts to break detailed balance and leads 
to closed coupled equations for a correlator and its conjugated
response function.
As mentioned above, these equations can be alternatively derived
from the driven Langevin dynamics of the $p$-spin model~\cite{leticia2}.
These dynamic equations were then
solved in the plane (Driving force, Temperature) in the case of a
constant driving force, and the results interpreted in the language
of nonlinear rheology, the driving force being analogous
to a shear stress $\sigma$.
Recall that in the absence of the shearing force, 
the model has a dynamic transition at a temperature $T_c$ where
the relaxation time diverges as a power law. Just above $T_c$, 
one has the standard two-step decay of the correlation, characterized 
by functional forms described in detail in Ref.~\cite{gotze}.

\begin{figure}
\begin{center}
\psfig{file=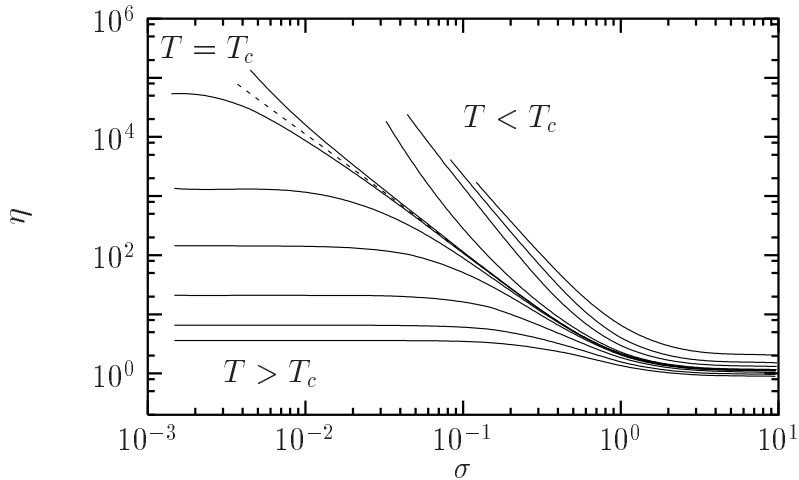,height=6.5cm}
\end{center}
\caption{\label{flowcurve} Flow curves $\eta = \eta(\sigma)$
for various temperatures above, at, and below $T_c$,
in the model of Ref.~\cite{BBK}.
Newtonian behavior is obtained at low $\sigma$, high $T$. 
Power law shear-thinning
is obtained everywhere else, with $\eta \propto \gamma^{-2/3}$ at $T_c$.}
\end{figure}

In the presence of the driving force, 
the whole physical behavior is encoded in the time decay of 
correlation and response functions. Their analysis leads to predictions
at several levels~\cite{BBK}. 
At the macroscopic level first, one gets flow curves relating the 
viscosity $\eta \equiv \int dt C(t)$ to the
shear stress $\sigma$, see Fig.~\ref{flowcurve}.
Beyond a linear regime restricted to the high-$T$, low-$\sigma$ region, 
the system is strongly shear-thinning. The power law 
$\eta \propto \gamma^{-2/3}$ is obtained at $T = T_c$, 
where $\gamma = \sigma /\eta$ is the shear rate.
In the glassy phase, $T<T_c$, the model describes a power law fluid, 
$\eta \approx \gamma^{-\alpha(T)}$, with no 
`dynamic' yield stress ($\alpha(T) <1 \Rightarrow \lim_{\gamma \to 0}
\sigma(\gamma) = 0$)~\cite{yield}.
Recall that the shear-thinning behavior is {\it derived} 
in the present framework without any assumption. 
The predicted macroscopic behavior compares
very well with experiments~\cite{exprheo} 
and simulations~\cite{onuki,BB1}.

Second, in the spirit of MCT, predictions can be made concerning the 
functional form of the correlators in various time regimes. 
They are the so-called `factorization property' at intermediate times
and the `superposition principle' at large times.
Again, these are well verified in simulation works~\cite{BB1}. 
We are not aware
of any experimental checks for the moment.

Third, a modified fluctuation-dissipation relation between 
correlation and response is also derived. It reads
\begin{equation}
\chi(t) = - \frac{1}{T_{\rm eff}} \frac{dC(t)}{dt},
\end{equation}
where the effective temperature $T_{\rm eff}$
replaces the equilibrium temperature $T$~\cite{leticia3,Cuku}. 
While the short processes show $T_{\rm eff} = T$, at long time
one gets $T_{\rm eff} > T$, showing that slow processes 
are `quasi-equilibrated' at an effective temperature higher than the thermal
bath temperature.
Again, extensive numerical simulations have confirmed this prediction
for various physical observables in a sheared fluid~\cite{BB1,BB2,BB3,liu}. 
Again, this important prediction has not been checked experimentally in this
context, although specific protocols have been described 
in Refs.~\cite{BB1,BB3}. 
We refer to Ref.~\cite{Teffexp} for experimental studies of aging systems.

\section{Yield stress and activated processes}
\label{yield}

Having access to a microscopic (spin glass)
model behind the  dynamical equations 
allows us to understand the evolution from 
the geometry of  the corresponding phase space. 
At zero external drive, this connection has been studied in much 
detail~\cite{Kupavi,Crso,Anire,ABarrat}.
Above the dynamical transition temperature $T_c$,
the available phase space is dominated by one large
basin in the free energy, corresponding to the `paramagnetic', 
or `liquid', state. 
At $T_c$, a threshold level in free energy appears, below
which the  free energy surface is split into  
exponentially-many disconnected regions. 

The aging dynamics below $T_c$ can be understood as a
gradual descent to the threshold level, starting from high energy
configurations~\cite{Cuku}. The slowing down is the
consequence of the decreasing connectivity of the visited landscape. 
See Ref.~\cite{andrea} 
for recent similar statements at equilibrium.
When the system is quenched from a high temperature, but at the same
time driven by non-conservative forces, it  remains 
similarly drifting  above
the free energy threshold, constantly receiving energy from the drive. 

On the other hand, if the system is prepared in one of the
deep regions below the threshold,
it remains trapped for all times~\cite{ABarrat}.
From this picture, we expect in that case that a weak
driving force will have essentially no effect beyond a
trivial `elastic' response of the system, 
as it is not strong enough to make the
system overcome the barriers. If instead a strong drive is applied, the
system should escape the low-lying valley and surfaces above the
threshold, where the  drive will suffice to keep it forever.

\begin{figure}
\begin{center}
\psfig{file=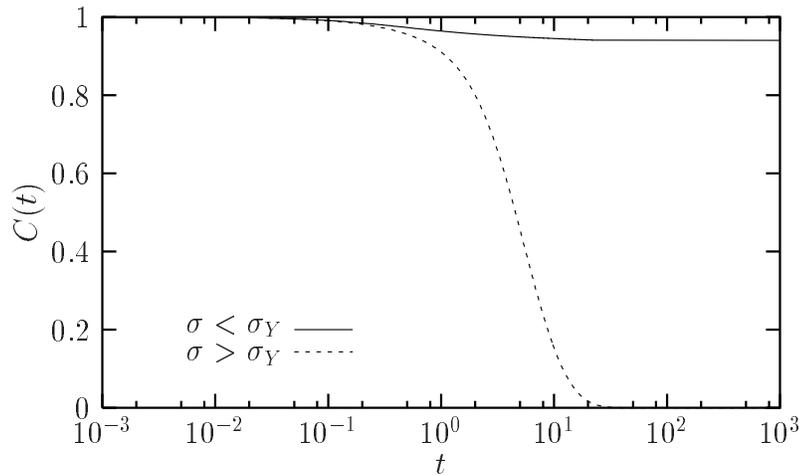,height=6.5cm}
\end{center}
\caption{\label{yieldstress} Existence of a 
static yield stress. The system 
is first prepared below the free energy threshold, as in Ref.~\cite{ABarrat}.
The driving force is then applied. The two 
curves are both taken in the subsequent nonequilibrium steady state
and show solid, $\sigma < \sigma_Y$, or liquid, $\sigma > \sigma_Y$, 
behaviors; $T=0.1$.}
\end{figure}

We have investigated this situation 
using the method of Refs.~\cite{leticia2,silvio}. A typical result
is shown in Fig.~\ref{yieldstress}, which shows that the above 
expectations deduced from the topology of the free energy landscape
are well verified when the calculations are done.
This proves the existence
of a {\it static yield stress} $\sigma_Y(T)$ in the model~\cite{yield}.
Figure~\ref{yieldstress} 
is also clearly reminiscent of the thixotropic 
behaviors~\cite{thixo} commonly encountered in soft glassy materials.

The difference between the free energy threshold and the 
equilibrium energy vanishes for $T \to T_c^-$, and the very notion
of a threshold disappears.
One expects therefore,
on physical grounds, that $\lim_{T \to T_c^-} \sigma_Y(T) = 0$.
This expectation is in qualitative disagreement with
Ref.~\cite{fuchs}. We are currently studying
the temperature dependence of $\sigma_Y(T)$ in our model to 
conclude on this issue.

We are now in a position to recall what is the main shortcoming 
of this mode-coupling type of approach. In any  
realistic system the structure of threshold and
valleys may remain essentially the same, but now {\it activated
processes} allow to jump barriers that are impenetrable at the 
perturbative level. In the spin language, barriers between states diverge
at the thermodynamic limit, $N\to\infty$, where $N$ is the total number
of spins. 

\begin{figure}
\hspace*{-1.2cm}
\psfig{file=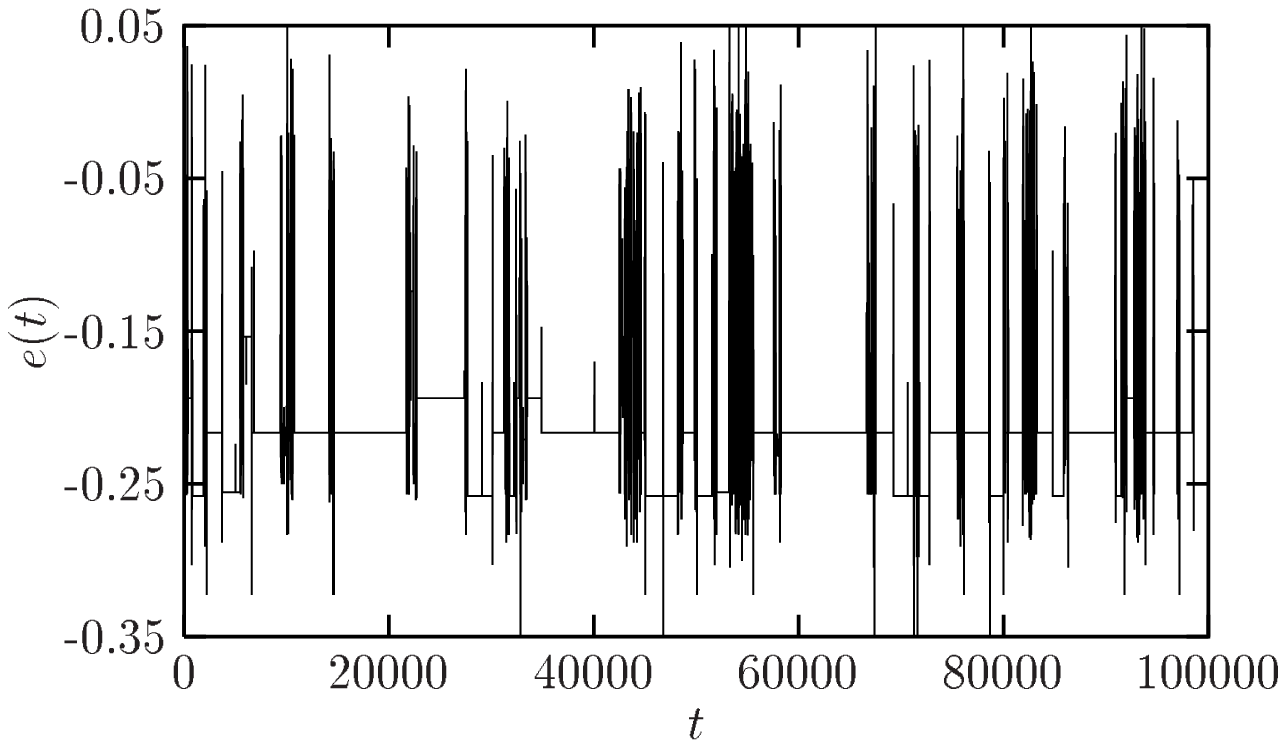,height=4.3cm} 
\psfig{file=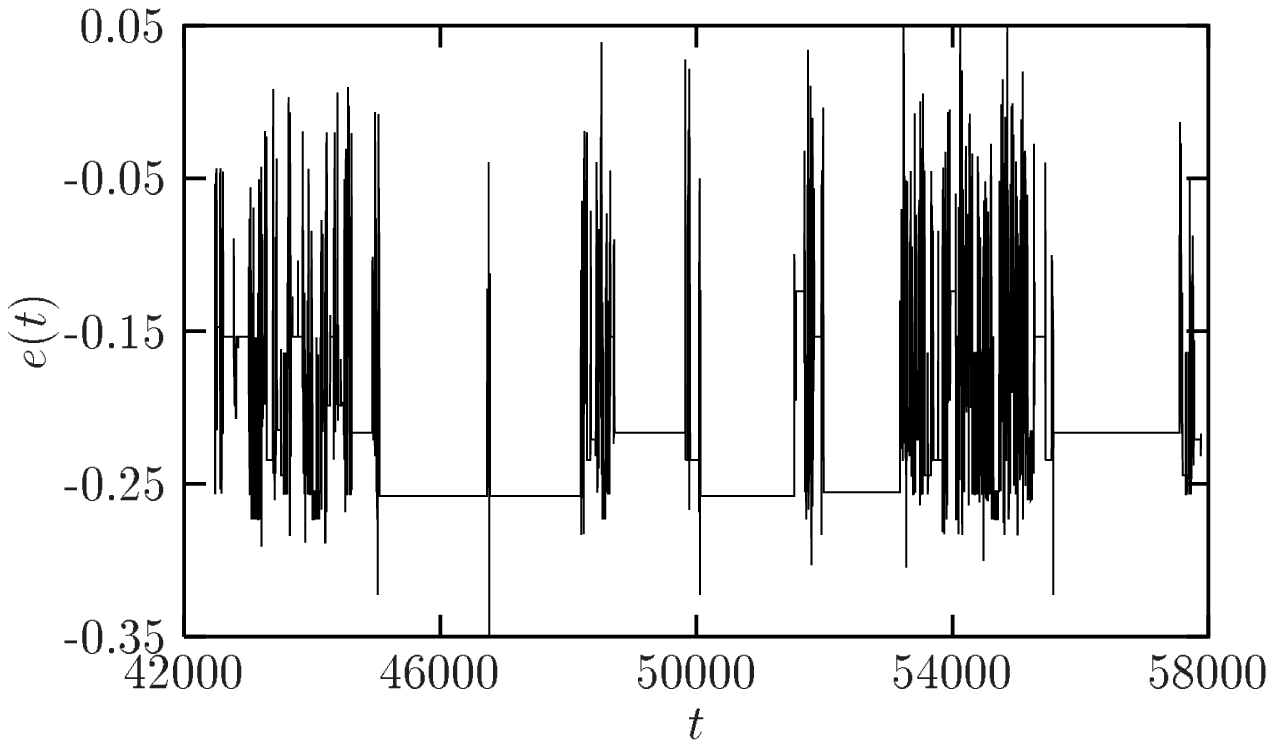,height=4.3cm} 
\hspace*{-1.2cm}
\psfig{file=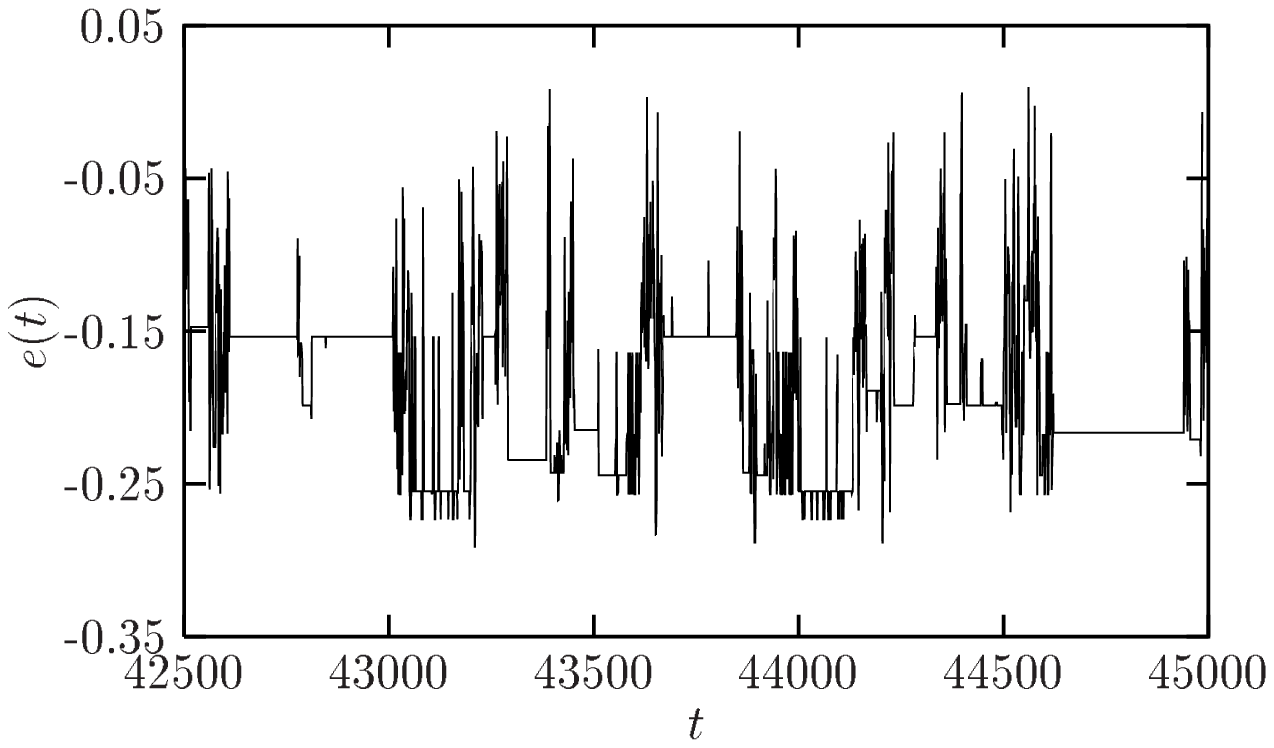,height=4.3cm} 
\psfig{file=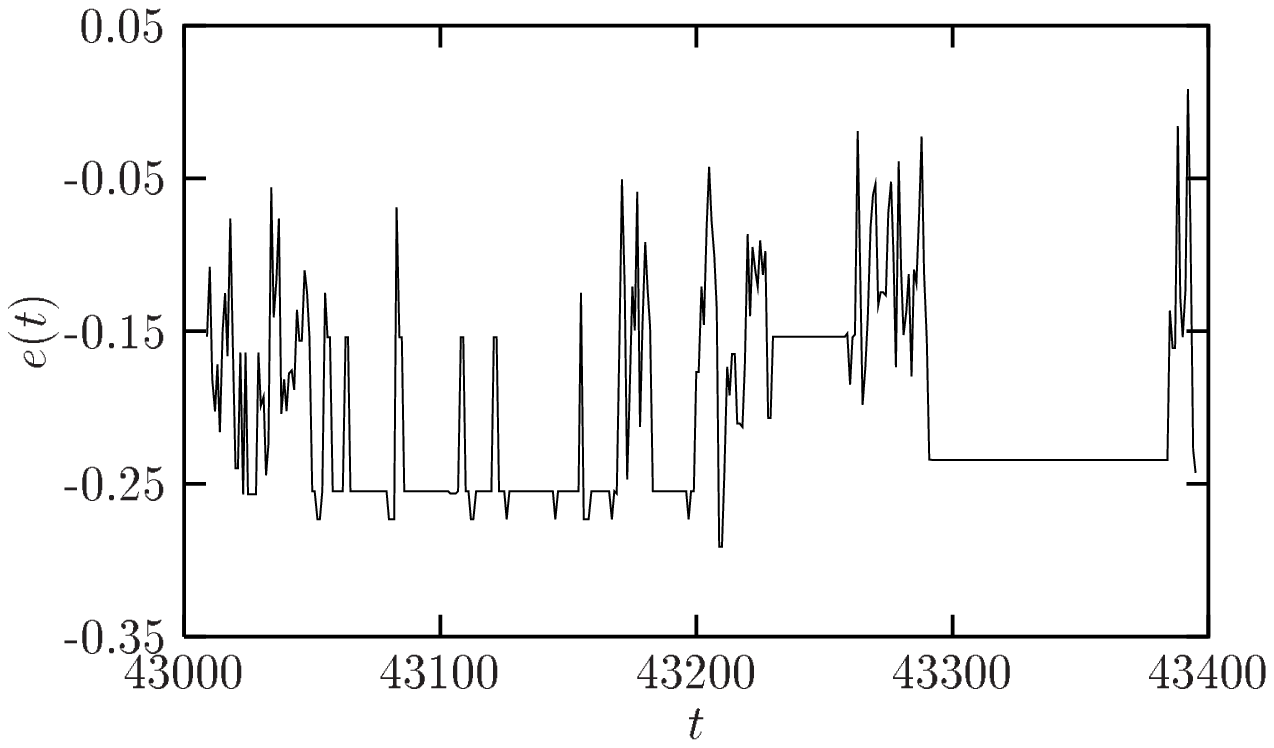,height=4.3cm} 
\caption{\label{self1} Self-similar
activated dynamics within the metastable states
of the finite-$N$ version of the model driven by an external force.
The first curve (top left) is taken in a single realization of the process.
The 3 other curves are successive zooms of the first one, as can be seen from
the time axis. A magnification factor of $\approx$ 6 is 
applied 3 times until elementary jumps are resolved in the last figure
(bottom right). Parameters are $N=20$, $T=0.05$.}
\end{figure}

Despite several attempts, the analytical inclusion of activation 
in this framework remains elusive.
However, a qualitative understanding may be gained by studying
the driven dynamics 
of the finite-$N$ version of the spin glass model
corresponding to our theory.
Keeping $N$ finite directly implies that barriers are finite, 
allowing thermal activation to play the role
it cannot play in the thermodynamic limit~\cite{leticia2}.
This type of study would not be possible within the MCT framework
alone. We have thus investigated by means of Monte Carlo simulations the
Ising $p$-spin model (for $p=3$), defined by the Hamiltonian
\begin{equation}
H = - \sum_{i < j < k}^N J_{ijk} s_i s_j s_k, 
\end{equation}
where $J_{ijk}$ is a random Gaussian variable of mean 0 and variance
$\sqrt{3}/N$. 
The driving force is implemented by the use of 
{\it asymetric} coupling constant, the amplitude of which being the
driving force, which we call $\sigma$, since it has the same
role as the shear stress in Ref.~\cite{BBK}; 
see also~\cite{leticia2,leticia3}. 
We numerically find clear evidences for the existence 
of a critical driving force $\sigma_Y(T)$ 
below which the system is trapped 
(`solid'), and above which it is not (`liquid'). 
The novelty is that activated processes now play an important
role. This can be observed in Fig.~\ref{self1} where the 
time dependence of the energy density 
is represented for a driving force amplitude
just below the yield value.
Strong fluctuations are observed in the energy density, and the system
alternates via thermal activation
between trapping periods and 
periods of freedom~\cite{leticia2,BCI}.

\begin{figure}
\begin{center}
\psfig{file=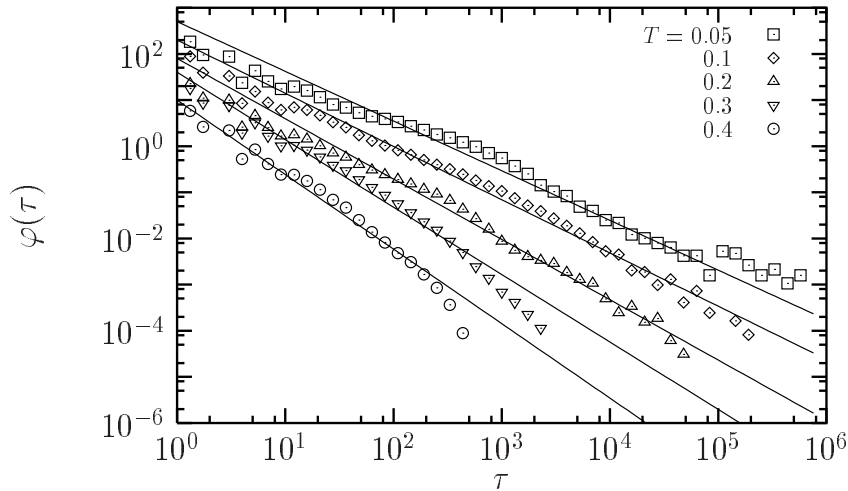,height=6.5cm}
\end{center}
\caption{\label{self2}
Distribution of trapping times $\varphi(\tau)$ for various temperatures 
in the presence of activation. The different curves are vertically shifted
for clarity.
The full lines are the theoretical prediction,
$\varphi(\tau) \propto \tau^{-(1+T/T_c)}$, with no fitting parameter.}
\end{figure}

Interestingly, Fig.~\ref{self1} also shows that this time evolution
is self-similar, in the sense that zooming on a particular time window 
leads to a very similar picture. In Fig.~\ref{self1},
a zooming procedure by a factor of order 6 is repeated 3 times
until unitary moves are resolved in the last figure.
This strongly 
suggests that trapping times are
power-law distributed. 
This is indeed 
what we find when the pictures of Fig.~\ref{self1} are 
quantitatively analyzed to extract 
the distribution of trapping times $\varphi(\tau)$.
This is shown in Fig.~\ref{self2}, together with excellent
comparison
to the theoretical prediction $\varphi(\tau) \sim \tau^{-\beta(T)}$
with an exponent $\beta(T) = 1+T/T_c$ which is discussed below.
Remark that the mean trapping time $\langle \tau
\rangle = \int d\tau' \varphi(\tau') \tau'$ is infinite
for $\beta(T)<2$, which
is equivalent to say that $\sigma < \sigma_Y$.
Consistently with the previous discussion, we find that $\sigma_Y(T)$
decreases when $T$ is increased, and our preliminary
results indicate that $\sigma_Y(T \to T_c^-)\to 0$.
More quantitative studies are also in progress.

This self-similar behavior is theoretically expected since a power law 
distribution can be obtained invoking activation dynamics 
in a landscape with exponentially distributed energy barriers, 
as is the case of the random energy model~\cite{REM}, 
obtained in the $p \to \infty$ limit.
This view is obviously reminiscent of the dynamics 
of the trap model~\cite{trapmodel},
the connection between the two approaches
being mathematically described in Ref.~\cite{benarous}.
These considerations in fact led us to predict
the temperature dependence of the 
trapping times distribution in Fig.~\ref{self2}
where we took $\beta(T) = 1+T/T_c$.
Note also the similarity between Fig.~\ref{self1} and recent 
numerical investigations of the `potential energy landscape'
at equilibrium around $T_c$, where
results were quantitatively
interpreted in the framework of the trap model~\cite{traps}.

\section{Flow heterogeneities}
\label{flow}

\begin{figure}
\begin{center}
\psfig{file=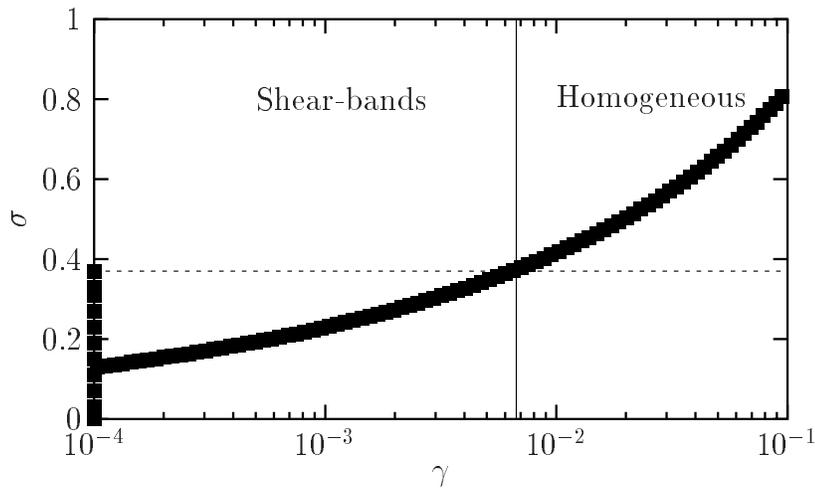,height=6.5cm}
\end{center}
\caption{\label{band1} A singular, or `nonmonotonic', flow curve
for $T=0.1$ in the model of Ref.~\cite{BBK}.
The horizontal dashed line is the value of the yield stress at this 
temperature.
The region where spatial
flow heterogeneities (shear-bands) may appear is delimited with the vertical
line at $\gamma=\gamma_c$ (see text).
Vertical points symbolizes the fact that any shear stress 
$\sigma < \sigma_Y$
is possible 
when $\gamma=0$ (this value of $\gamma$
is of course not visible in a log-scale).}
\end{figure}

The results presented in
the two previous sections have interesting implications that
we now discuss in detail.
We noted in section~\ref{steady} that our model 
describes a power law fluid at low temperatures, $\sigma \approx 
\gamma^{1-\alpha(T)}$, the shear-thinning exponent satisfying
$\frac{2}{3} = \alpha(T_c) < \alpha(T) < 1$.
When the consideration on the yield stress
of section~\ref{yield} are included, the complete 
flow curve of the material admits a singular $\gamma \to 0$ limit, see
Fig.~\ref{band1}.
Hence, the flow curve becomes `nonmonotonic'. 
This singularity is well-known at the experimental 
level~\cite{coussot,yield}.
Note that it physically results, in our theoretical
description, from the existence of the threshold value in the 
free energy. 
Note also that a purely dynamical approach \`a la MCT
misses this subtetly~\cite{fuchs}.

\begin{figure}
\begin{center}
\psfig{file=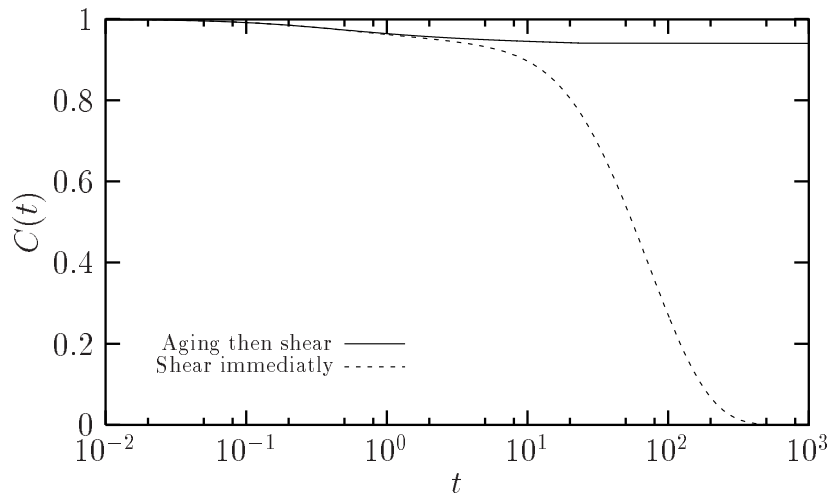,height=6.5cm}
\end{center}
\caption{\label{band2} Two possible dynamical solutions 
for $T=0.1$ and $\sigma=0.1$ in the model of Ref.~\cite{BBK}.
They coexist in a single sample when the
system displays shear-bands.
Compare with Fig.~2 of Ref.~\cite{simubands}.}
\end{figure}

A direct consequence is that there is a complete interval of shear stresses, 
$\sigma \in [0:\sigma_Y]$, for which two dynamical solutions are 
possible. 
This is examplified in Fig.~\ref{band2}, where the correlation function
is represented for the same value of the shear stress, but starting
either from random initial conditions (`shear immediatly'), 
or from an equilibrated initial condition below
the free energy threshold (`aging then shear').
Both curves are obtained in a driven steady state for the {\it same}
value of the control parameters $(T,\sigma)$, and are equally stable 
at the mean-field level.

Looking again at Fig.~\ref{band1}, one sees 
that the flow curve defines a critical
value $\gamma_c$ of the shear rate, 
defined by $\sigma(\gamma_c) = \sigma_Y$
and represented by a vertical line in the figure.
A very interesting question now is:
what happens in an experiment if a {\it global}
shear rate $\gamma_{\rm global} < \gamma_c$ is applied to the sample?
An homogeous flow would indeed correspond to a stress
$\sigma < \sigma_Y$ for which a second dynamical solution exists.
The answer is known from experiments~\cite{expbands} 
and simulations~\cite{simubands}: the system
will spontaneously develop {\it flow heterogeneities}, where
a flowing band ($\gamma_{\rm local} > 0$) coexist with a non-flowing
band ($\gamma_{\rm local}=0$), both supporting the same value of the shear
stress.
In that case, flowing regions will display a dynamic behavior
described by the dashed line in Fig.~\ref{band2}, whereas the second,
jammed, band
will be described by the full line in Fig.~\ref{band2}.
This is also observed in a recent numerical simulation, see Fig.~2 of 
Ref.~\cite{simubands}.
This shear-banding phenomenon results then
from the `nonmonotonicity' of the flow curve in Fig.~\ref{band1},
and can again 
ultimately be viewed, in our model, as an experimental consequence 
of the notion of threshold in the free energy landscape.

An open problem, of course, is the dynamical selection of the bands
in a real sheared material~\cite{guillemette,dhont}. 
How does the system choose the 
relative size of the bands?
The problem is difficult, since no thermodynamic argument (such as 
a Maxwell construction) can be applied to this nonequilibrium situation. 
The same question is presently 
much discussed in various complex fluids, like liquid crystals~\cite{liquid}
or wormlike micelles~\cite{suzanne}.

\section{Conclusion}
In these proceedings, preliminary results concerning the 
theoretical description of the yield stress, the flow heterogeneties
and the role of activated processes in soft glassy materials 
via a nonequilibrium schematic mode-coupling theory were presented. 
An advantage of our approach is that no assumption are made,
at odds with the more phenomenological models which are usually
used in the field of rheology.
This allows to make detailed microscopic predictions, 
much beyond the macroscopic rheological level
where concurrent models are stuck. These predictions were briefly
reviewed in section~\ref{steady}.
We emphasize, as we did in Ref.~\cite{BB3}, that many of them
are still experimentally unverified. In that sense, the situation
is very similar to the mid-1980's, where schematic mode-coupling models
were derived, but with little experimental confirmation
of their main features.

We mentioned at several places of this paper that
our results were only preliminary, and we took advantage of this conference
to present our `work in progress'.  
Although more precise and complete studies are obviously 
necessary, we have shown that a qualitative understanding 
of the rheological behavior could be gained via this approach,
beyond the steady rheological situation
studied in Ref.~\cite{BBK}.

A more quantitative approach would be to derive a `full' nonequilibrium
theory for sheared fluids. This program was recently undertaken
by Fuchs and Cates~\cite{fuchs}.
Their derivation makes use of a projector
formalism instead
of the perturbative development described in section~\ref{approach}, and
ends up with dynamic equations for correlators only.
It is not clear at the moment if the qualitative discrepancies 
between the two approaches underlined in this paper
are due to this alternative derivation where 
nonequilibrium effects (like the violation of the fluctuation-dissipation
theorem) are not described.
Comparisons between the two approaches would thus be very interesting.

Last, we mention also that if flow heterogeneities are suggested 
in our model by the existence of two dynamical solutions for a given
shear stress, this does not imply that shear-banding has to 
actually take place,
nor does it allow to gain any insight into 
the problem of the selection between the two solutions. 
This certainly requires more ambitious approaches than ours. 

\ack

This paper is dedicated to $2\varepsilon$ who became $O(1)$ 
during this conference.

\vspace*{3mm}

\noindent
This work was motivated by, and benefited from interactions with
A.~Ajdari, J.-L.~Barrat, G.~Biroli, 
J.-P.~Bouchaud, M.~Cates, L.~F.~Cugliandolo, 
J.~Kurchan, A.~Lef\`evre, and G.~Picard during 
the ``Slow Relaxations and nonequilibrium dynamics 
in condensed matter" Les Houches Summer School, July 2002.
Discussions with L.~Bocquet, M.~Fuchs, J.~P.~Garrahan,
D.~Sherrington, and F.~Varnik are also acknowledged.
G.~T\'oth is thanked for his absence during the preparation
of the manuscript.
Numerical results were obtained on OSWELL
at the Oxford Supercomputing Center, Oxford University, UK.
Our work is supported by the EPSRC (UK).


\section*{References}

\begin{thebibliography}{99}

\bibitem{sollich1} P. Sollich, F. Lequeux, P. H\'ebraud, and M.E. Cates, 
Phys. Rev. Lett. {\bf 78}, 2020 (1997).

\bibitem{trapmodel} J.-P. Bouchaud, J. Phys. I (France) {\bf 2}, 1705 (1992).

\bibitem{sollich2} P. Sollich, Phys. Rev. E {\bf 58}, 738 (1998);
S. M. Fielding, P. Sollich, and M. E. Cates, J. Rheol {\bf 44}, 323 (2000).

\bibitem{derec} C. D\'erec, A. Ajdari, G. Ducouret, and F. Lequeux, 
C. R. Acad. Sci. Paris IV {\bf 1}, 1115 (2000);
C. D\'erec, A. Ajdari, and F. Lequeux, Eur. Phys. J. E {\bf 4}, 355 (2001).

\bibitem{lemaitre} A. Lemaitre, preprint cond-mat/0206417.

\bibitem{coussot} P. Coussot, Q. D. Nguyen, H. T. Huynh, and D. Bonn, 
J. Rheol {\bf 46}, 573 (2002).

\bibitem{BBK} L. Berthier, J.-L. Barrat, and J. Kurchan,
Phys. Rev. E {\bf 61}, 5464 (2000).

\bibitem{fuchs} M. Fuchs and M. E. Cates, preprint cond-mat/0204628,  
cond-mat/0207530.

\bibitem{bouchaud} J.-P. Bouchaud, L.F. Cugliandolo, J. Kurchan, and 
M. M\'ezard, Physica A {\bf 226}, 243 (1996).

\bibitem{gotze} W. G\"otze, in {\it Liquids, Freezing and Glass Transition}, 
Eds.: J.P. Hansen, D. Levesque and J. Zinn-Justin, Les Houches
 1989, (North Holland, Amsterdam, 1991); W. G\"otze and L.  Sj\"ogren,
Rep. Prog. Phys. {\bf 55}, 241 (1992); W. G\"otze,
J. Phys. Condens. Matter {\bf 11}, A1 (1999).

\bibitem{mazenko} S. P. Das, G. F. Mazenko, S. Ramaswamy, and J. J. Toner,
Phys. Rev. Lett. {\bf 54}, 118 (1985).

\bibitem{kawasaki} K. Kawasaki and S. Miyazima, 
Z. Phys. B {\bf 103}, 423 (1997).

\bibitem{schematic} U. Bengtzelius, W. G\"otze, and A. Sj\"olander,
J. Phys. C: Solid State Phys. {\bf 17}, 5915 (1984);
E. Leutheusser, Phys. Rev. A {\bf 29}, 2765 (1984).

\bibitem{tarjus} G. Tarjus and D. Kivelson, in
{\it Jamming and Rheology: Constrained dynamics on microscopic and
macroscopic scales}, Eds.: A. Liu and S. R. Nagel (Taylor and Francis,
New York, 2001).

\bibitem{reviewaging} J.-P. Bouchaud, L.F.  Cugliandolo, 
J. Kurchan, and M. M\'ezard in {\it
Spin Glasses and Random Fields}, Ed.: A.P. Young (World Scientific,
Singapore, 1998); preprint cond-mat/9511042.

\bibitem{kithprl} T. R. Kirkpatrick and D. Thirumalai, 
Phys. Rev. Lett. {\bf 58}, 2091 (1987).

\bibitem{series} T. R. Kirkpatrick and D. Thirumalai,
Phys. Rev. B {\bf 36}, 5388 (1987);
T. R. Kirkpatrick and P. G. Wolynes,
Phys. Rev. A {\bf 35}, 3072 (1987).

\bibitem{leticia2} L.F. Cugliandolo, J. Kurchan, P. Le Doussal,
and L. Peliti, Phys. Rev. Lett. {\bf 78}, 350 (1997).

\bibitem{yield} H. A. Barnes, J. Non-Newtonian Mech. {\bf 81}, 133 (1999).

\bibitem{exprheo} D. Bonn, S. Tanase, B. Abou, H. Tanaka, and 
J. Meunier, Phys. Rev. Lett. {\bf 89}, 015701 (2002);
S. Bair, 
J. Non-Newtonian Fluid Mech. {\bf 97}, 53 (2001);
J. D. Ferry,
{\it Viscoelastic properties of polymers} (Wiley, New York, 1980);
R. G. Larson, {\it The structure and rheology 
of complex fluids} (Oxford University Press, Oxford, 1999).

\bibitem{onuki}
R. Yamamoto and A. Onuki, Phys. Rev. E {\bf 58}, 3515 (1998).

\bibitem{BB1}
L. Berthier and J.-L. Barrat,  
J. Chem. Phys. {\bf 116}, 6228 (2002). 

\bibitem{leticia3} L.F. Cugliandolo, J. Kurchan, and L. Peliti, Phys. 
Rev. E {\bf 55}, 3898 (1997).

\bibitem{Cuku} L.F. Cugliandolo and J. Kurchan,
Phys. Rev. Lett. {\bf 71}, 173 (1993).

\bibitem{BB2} J.-L. Barrat and L. Berthier,
Phys. Rev. E {\bf 63}, 012503 (2001).

\bibitem{BB3}
L. Berthier and J.-L. Barrat, Phys. Rev. Lett. {\bf 89}, 095702 (2002).

\bibitem{liu} I. K. Ono, C. S. O'Hern, D. J. Durian, S. A. Langer, 
A. J. Liu, and S. R. Nagel, 
Phys. Rev. Lett. {\bf 89}, 095703 (2002).

\bibitem{Teffexp} T. S. Grigera and N. E. Israeloff, Phys. Rev. Lett. 
{\bf 83}, 5038 (1999);
L. Bellon, S. Ciliberto, and C. Laroche, Europhys. Lett. {\bf 53}, 
511 (2001); D. H\'erisson and M. Ocio,
Phys. Rev. Lett. {\bf 88}, 257202 (2002).

\bibitem {Kupavi} J. Kurchan, G. Parisi, and M.A. Virasoro, J. Phys. I France 
{\bf 3}, 1819 (1993).

\bibitem{Crso} A. Crisanti and H.J. Sommers, J. Phys. I France 
{\bf 5}, 805 (1995). 

\bibitem{Anire} A. Cavagna, I. Giardina, and G. Parisi, Phys. Rev. B {\bf 57},
11251 (1998).

\bibitem{ABarrat} A. Barrat, R. Burioni, and M. M\'ezard, J. Phys. A {\bf 29},
 L81 (1996).

\bibitem{andrea}
K. Broderix, K. K. Bhattacharya, A. Cavagna, A. Zippelius, and I. Giardina,
Phys. Rev. Lett. {\bf 85}, 5360 (2000);
L. Angelani, R. Di Leonardo, G. Ruocco, A. Scala, and
F. Sciortino, Phys. Rev. Lett. {\bf 85}, 5356 (2000).

\bibitem{silvio} S. Franz and G. Parisi,
J. Phys. I (France) {\bf 5}, 1401 (1995)

\bibitem{thixo} H. A. Barnes, J. Non-Newtonian Mech. {\bf 70}, 1 (1997).

\bibitem{BCI} L. Berthier, L. F. Cugliandolo, and J. L. Iguain, 
Phys. Rev. E {\bf 63}, 051302 (2001).

\bibitem{REM} B. Derrida, Phys. Rev. Lett. {\bf 45}, 79 (1980);
Phys. Rev. B {\bf 24}, 2613 (1981). 

\bibitem{benarous} G. Ben Arous, A. Bovier, and V. Gayrard, 
Phys. Rev. Lett. {\bf 88}, 087201 (2002).

\bibitem{traps} R. A. Denny, D. R. Reichman, and J.-P. Bouchaud,
cond-mat/0209020; B. Doliwa and A. Heuer, preprint cond-mat/0205283.

\bibitem{expbands}
R. Mas and A. Magnin, J. Rheol. {\bf 38}, 889 (1994);
J. Persello, A. Magnin, J. Chang, J. M. Piau, and B. Cabane, 
J. Rheol. {\bf 38}, 1845 (1994);
F. Pignon, A. Magnin, and J.-M. Piau, J. Rheol., {\bf 40}, 573 (1996);
G. Debr\'egeas, H. Tabuteau, and J.-M. di Meglio,
Phys. Rev. Lett. {\bf 87}, 178305 (2001);
R. Mas and A. Magnin, J. Rheol. {\bf 38}, 889 (1994);
P. Coussot, J. S. Raynaud, F. Bertrand, P. Moucheront, 
J. P. Guilbaud, H. T. Huynh, S. Jarny, and D. Lesueur, 
Phys. Rev. Lett. {\bf 88}, 218301 (2002).

\bibitem{simubands}  F. Varnik, L. Bocquet, J.-L. Barrat, and L. Berthier,
preprint cond-mat/0208485.

\bibitem{guillemette}
G. Picard, A. Ajdari, L. Bocquet, and F. Lequeux,
preprint cond-mat/0206260.

\bibitem{dhont} J. K. G. Dhont, 
Phys. Rev. E {\bf 60}, 4534 (1999).

\bibitem{liquid} P. D. Olmsted and C. Y. D. Lu,
Phys. Rev. E {\bf 60}, 4397 (1999).

\bibitem{suzanne} S. M. Fielding and P. D. Olmsted,
preprint cond-mat/0208599.

\end{thebibliography}
\end{document}